\documentclass[10pt]{article}
\usepackage{amsmath}
\usepackage{amssymb}
\usepackage{graphicx}
\usepackage{natbib}

\def\aa{A\&A }

\def\aj{AJ }

\def\mnras{MNRAS }
\newcommand{\uplet}[2]{{#1}^{\mbox{\footnotesize\sc\hspace{-0.1em} #2}}}
\newcommand{\Frac}[2]{\frac{\displaystyle\strut #1}{\displaystyle\strut #2} }
\newcommand{\hem}{\hspace{1em}}

\begin{document}

\vspace*{0.5cm}

\noindent {\Large ON OBSERVABILITY OF THE FREE CORE NUTATION }
\vspace*{0.7cm}

\noindent\hspace*{1.5cm} L. PETROV \\
\noindent\hspace*{1.5cm} NVI, Inc./NASA GSFC \\
\noindent\hspace*{1.5cm} Code 698, NASA GSFC, Greenbelt, MD 20771 USA \\
\noindent\hspace*{1.5cm} Leonid.Petrov@lpetrov.net \\

\vspace*{0.5cm}

\noindent {\large ABSTRACT.} 
\vspace*{0.2cm}

  Neither astronomical technique, including VLBI, can measure nutation directly. 
Estimates of parameters of the nutation model are produced by solving 
the LSQ problem of adjusting millions parameters using estimates of group 
delay. The choice of the mathematical model for nutation used in the 
estimation process of analysis of group delays affects our ability to 
interpret the results. Ignoring these subtleties and using parameters of 
the nutation model either in the form of time series, or in the form of 
empirical expansion as "VLBI measurement of nutation", opens a room 
for misinterpretation and mistakes. Detailed analysis of the problem 
reveals that the separation of forced nutations, atmospheric nutations, 
ocean nutations, and the retrograde free core nutation requires invoking some 
hypotheses, and beyond a specific level becomes uncertain. This sets 
a limit of our ability to make an inference about the free core nutation.

\vspace*{0.5cm}

\section{\large INTRODUCTION}

   There are several basic misconceptions concerning the theory of nutation.

{\sf Misconception~1}: {\it ``VLBI measures nutation''}. In fact, VLBI measures 
\ldots the thermal noise at receivers with a tiny admixture of the noise from 
an observed extragalactic source. VLBI is the technique for the evaluation of 
the spectrum of the cross-correlation function using records of thermal noise
synchronized by independent clocks. Using cross-spectrum, group delays can be
estimated. The theory of wave propagation describes the dependence of group 
delay on the motion of the emitter and receivers, the properties of propagation
media, and the phase fluctuations inside the electronic equipment. This 
dependence can be reduced to a parametric model, and parameters of this
model can be adjusted using all available estimates of group delays.

  Thus, nutation parameters are not measured, but estimated together with
more than a million other parameters. Unlike to direct measurements, these
estimates are heavily depends on a subjective choice of parameterization.
One cannot adjust only nutation parameters --- in that case the fit would
be very poor. For this reason, one should not interpret nutation parameters
alone, they have sense only as a part of the overall mathematical model
that describes the motion of the station network.

  {\sf Misconception~2}: {\it ``There exists a theory of forced nutation with 
accuracy comparable to observations''}.  An elaborate theory of the non-rigid 
Earth nutation was developed by \citet{r:Wahr80} in 1970-s. It explained 91\% 
of the deviation of the real Earth nutations from the absolute rigid body 
nutations. Numerical values of nutation expansion in the framework of this 
theory were computed using some integral quantities that depends on profiles 
of density and elasticity parameters inside the Earth. These profiles were 
derived from analysis of seismological data. However, the disagreement of 
Wahr's theory with observations is currently at a 500-$\sigma$ level. 

\par\vspace{-2ex}\par\noindent
\begin{quotation}\it
If it disagrees with experiment it is wrong. In that simple statement is 
the key to science. It does not make any difference how beautiful your 
guess is. It does not make any difference how smart you are, who made 
the guess, or what his name is --- if it disagrees with experiment it 
is wrong. That is all there is to it.
\end{quotation}
\par\vspace{-2ex}\par\noindent\cite{r:fey}, page~156.
\par\medskip\par

  Thus, we are compelled to acknowledge that the theory of nutation is wrong.
Numerous attempts to improve the theory were undertaken, but they all failed. 
In recognition of this failure, some authors resorted to fitting parameters 
of their theories to the adjustments of nutation angles from VLBI analysis 
--- just the quantities that the theory is supposed to predict. Certainly, 
by fitting a set of ad~hoc parameters, one can get a set of coefficients 
of nutation expansion that may have whatever small fit to nutation angle
adjustments. But this mathematical trick does not make the theory correct,
and this set of coefficients cannot be called theoretical, but should be 
called empirical. Empirical expansions were presented in papers of
\citet{r:Her86,r:Getino2001,r:shir01,r:mhb2000,r:kra06b}.

  The fact there is no precise theory of forced nutation, has important
consequences. First, fitting parameters cannot be interpreted in terms
consistent with the failed theory. Second, the residuals between the empirical
theory and adjustments to nutation angles derived from VLBI analysis of
group delay should not be interpreted as quantities with a specific physical
meaning. 

  {\sf Misconception~3}: {\it ``An elaborate theory of nutation is needed for 
practical applications''}. Nutation has two major constituents: forced nutation 
with the precisely known excitation exerted by the Moon and the Sun and the 
constituents excited by the re-distribution of oceanic and atmospheric masses.
The latter term is unpredictable in principle. The accuracy of determination
of the nutation expansion from analysis of VLBI group delays has passed the
level of atmospheric nutation contribution in 1990s. Therefore, even a 
precise theory of forced nutation would have been built, that theory would not
be able to predict nutation with the accuracy comparable with observations.
Similar to Chandler wobble, nutation parameters will have to be always 
determined from observations without any theory in mind.

\section{\large DETERMINATION OF THE FREE CORE NUTATION}

  We consider here that $N$ stations observe $K$ celestial physical bodies. 
It is assumed that each station is associated with a reference point. 
In the case of VLBI antennas, this is the point of projection of the moving 
axis to the fixed axis. Observing stations receive electromagnetic radiation 
emitted by celestial bodies, and each sample of the received signal 
is associated with a time stamp from a local frequency standard synchronized 
with the GPS time. Analysis of voltage and time stamps of received radiation 
eventually allows us to derive the differences in the photon propagation time 
from observed bodies to reference points of observing stations. These 
distances depend on relative positions of stations with respect to observed 
bodies. The instantaneous coordinate vector of station $i$, 
in the inertial coordinate system at a given moment of time 
$\uplet{\vec{r}}{c}_i(t)$ is represented as the sum of a rotation and 
translation applied to a vector $\uplet{\vec{r}}{t}_i(t)$ in the terrestrial 
coordinate system as
\begin{eqnarray}
   \uplet{\vec{r}}{c}_i(t) = \widehat{\mathstrut\cal M}_a(t) \,\, 
                             \uplet{\vec{r}}{t}_i \: + \:
                             \vec{q}_e(t) \times 
                             \uplet{\vec{r}}{t}_i \: + \:
                             \uplet{\vec{d}}{t}_i(t) \: + \:
                             \vec{T}(t) 
\label{e:e1}\end{eqnarray}
  where $ \widehat{\mathstrut\cal M}_a(t) $ is the a~priori rotation matrix, 
$\vec{q}_e(t)$ is the vector of small rotation, $\vec{T}(t)$ is the 
translational motion of the network of stations, and $\vec{d}_i(t)$ is 
a displacement vector of an individual station. Equations of photon propagation 
tie the instantaneous vector of site coordinates $\vec{r}_i(t)$ with vectors 
of observed physical bodies and their time derivatives. These relationships 
allow us to build a system of equations of conditions. Solving these 
equations, we can get estimates of expansion of the vector $\vec{q}_e(t)$ over 
some basis functions. 

  It has been demonstrated by \citet{r:pet07} that coefficients of expansions
of the vector $\vec{q}_e(t)$ over the Fourier and B-spline bases can be
found directly in a single LSQ solution that uses VLBI group delays estimates 
without resorting to intermediate time series. This means that we can estimate
directly the spectrum of the Earth orientation variations from estimates of
VLBI group delays. The portion of the spectrum of variations in $q_1, q_2$
considered as a complex process $q_1 \; + \; i q_2$ in the frequency range 
$[-1.5\Omega, -0.5\Omega]$, where $\Omega$ is the 
positive nominal frequency of the Earth rotation 
\mbox{$7.292115146706979 \cdot 10^{-5}$ rad\ s${}^{-1}$}, is called nutation.

  The nutation spectrum can be represented as a sum of two components: the 
forced nutation --- a rail of with very sharp peaks with exactly known 
frequency, and the retrograde free core nutation (RFCN) --- a band-limited 
continuous process with frequencies around the retrograde free core nutation
frequency \mbox{$-7.30901 \cdot 10^{-5}$ rad\ s${}^{-1}$}. Extensive 
discussion of nutation theory can be found in the monograph of 
\citet{r:moritz} and in the modern paper of \citet{r:kra06}.

\begin{figure}
   \begin{center} 
      \caption{Spectrum of the free core nutations and forced nutations.
               Circles denote the power of the rigid Earth nutations, disks
               denote the power of the estimated nutation spectrum from VLBI 
               group delays.}
       \label{f:f1}
      \par\noindent\vspace{-3ex}\par\hphantom{a}\par
      \includegraphics[width=80mm,height=60mm]{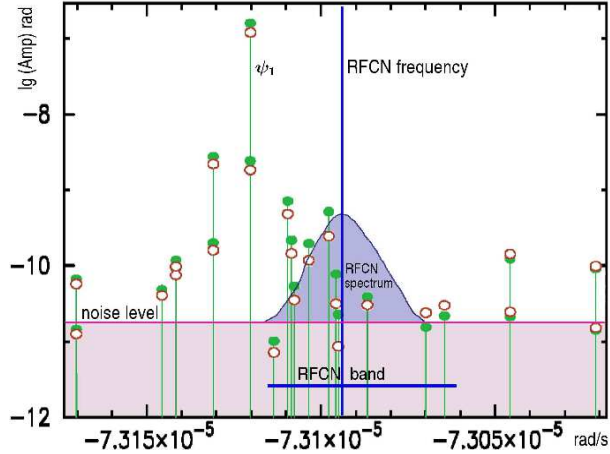}
   \end{center}
   \par\noindent\vspace{-9ex}\par\hphantom{a}\par
\end{figure}

  The problem of estimation of the free core nutation is reduced to the 
filtration of the observed spectrum of the Earth's rotation variations.
The principle difficulty in separation of the RFCN from forced nutations is 
that their spectrum is overlapping, as figure~\ref{f:f1} demonstrates.

  We can separate the two constituents assuming 1) the RFCN is a band-limited
process with known frequency band and unknown excitation; 2) the forced
nutation is the purely harmonic process with known frequencies, known
excitation, and a response function being {\it a smooth function} of frequency.

  For practical implementation of this approach one should determine a)~the
band of the RFCN process and b)~the empirical response function of forced 
nutations.

  Empirical response function can be found with a data mining technique 
by comparing the spectrum of the Earth orientation with nutation constituents
computed for the mechanical model of the absolutely rigid model. After 
a relatively short search, we find that complex amplitudes of forced nutations 
from VLBI data analysis $A_e$ can be very well approximated to the theoretical
amplitudes $A_r$ for the mechanical model of the absolutely rigid Earth through
this expression:
\begin{eqnarray}
   A_e(\omega) = [ \alpha(\omega - \beta) + \Frac{\gamma}{\omega - \delta} ] 
                 \, (A_r(\omega) - A_n(\omega))
\label{e:e2}\end{eqnarray}
   where $\omega$  is the frequency, $\alpha, \beta, \gamma$, and $\delta$
are complex parameters. $A_n$ is the complex nutation amplitude caused by
the non-tidal excitation, for instance, by the ocean and the atmosphere. 
It should be noted that functional dependence in the brackets of expression
of \ref{e:e2}, also called transfer function, can be derived analytically 
assuming 1)~the triaxiality of the Earth's inertia ellipsoid is negligible; 
2)~the Earth consists of an elastic mantle and a liquid core; 3)~the Earth's 
response is linear to the external torques. It should be stressed that 
although a simple theoretical considerations allow us to derive 
a parameterized expression similar to \ref{e:e2}, the theory predicts 
{\it wrong} numerical coefficients, and therefore, it cannot be trusted.

  Parameters $\alpha, \beta, \gamma$, and $\delta$ can be determined from 
analysis of VLBI estimation results. We should be aware that 
1)~the estimates of coefficients $\alpha, \beta, \gamma$, and $\delta$ 
heavily depend on a small set of constituents, including the constituent
that corresponds to the $\psi_1$ tide, which is within the RFCN band; 
2)~estimation of the parameters of transfer function is a non-linear 
problem; 3)~the oceanic and atmospheric contribution to the excitation
has a significant uncertainty.

  Using the estimates of the empirical transfer function, we can compute 
predicted forced nutations within the RFCN frequency band, and interpret
the residual spectrum as the empirical RFCN spectrum. Predicted forced 
nutations have errors not only due to statistical uncertainties in the 
coefficients $\alpha, \beta, \gamma, \delta$, but also due to uncertainties 
in $A_f$ and due to the errors of approximation in the expression \ref{e:e2}. 
The fact that the spectrum of the RFCN and forced nutation is overlapping 
has important consequences: 1) the estimates of the RFCN spectrum have 
the statistical uncertainties due to the noise in VLBI group delay and 
the constraint uncertainties due to constituents separation, 2) estimates of 
the RFCN spectrum are not unique, but depend on assumptions made for 
parameterization of the empirical transfer functions; 3) estimates of the 
RFCN spectrum depend on the contribution of the ocean and the atmosphere 
on the main constituents of forced nutations.

\begin{figure}
   \begin{center} 
      \caption{Estimates of the RFCN power spectrum from VLBI time delays
               from 1984 through 2007 and resonance constraints on forced 
               nutation. A)~Empirical transfer function from $\psi_1, K_1, 
               P_1, O_1$, etc was extrapolated to the entire RFCN band; 
               B)~Forced nutations within the RFCN band were ignored and 
               as a result they propagated to the estimates of the RFCN 
               spectrum.}
      \label{f:f2}
      \par\vspace{0.5ex}\par\noindent
      \small
      \hspace{0.03\textwidth} 
      A) \hspace{0.45\textwidth} B)  \hspace{0.41\textwidth}\hphantom{a}
      \par\vspace{0.5ex}\par\noindent 
      \includegraphics[width=0.48\textwidth]{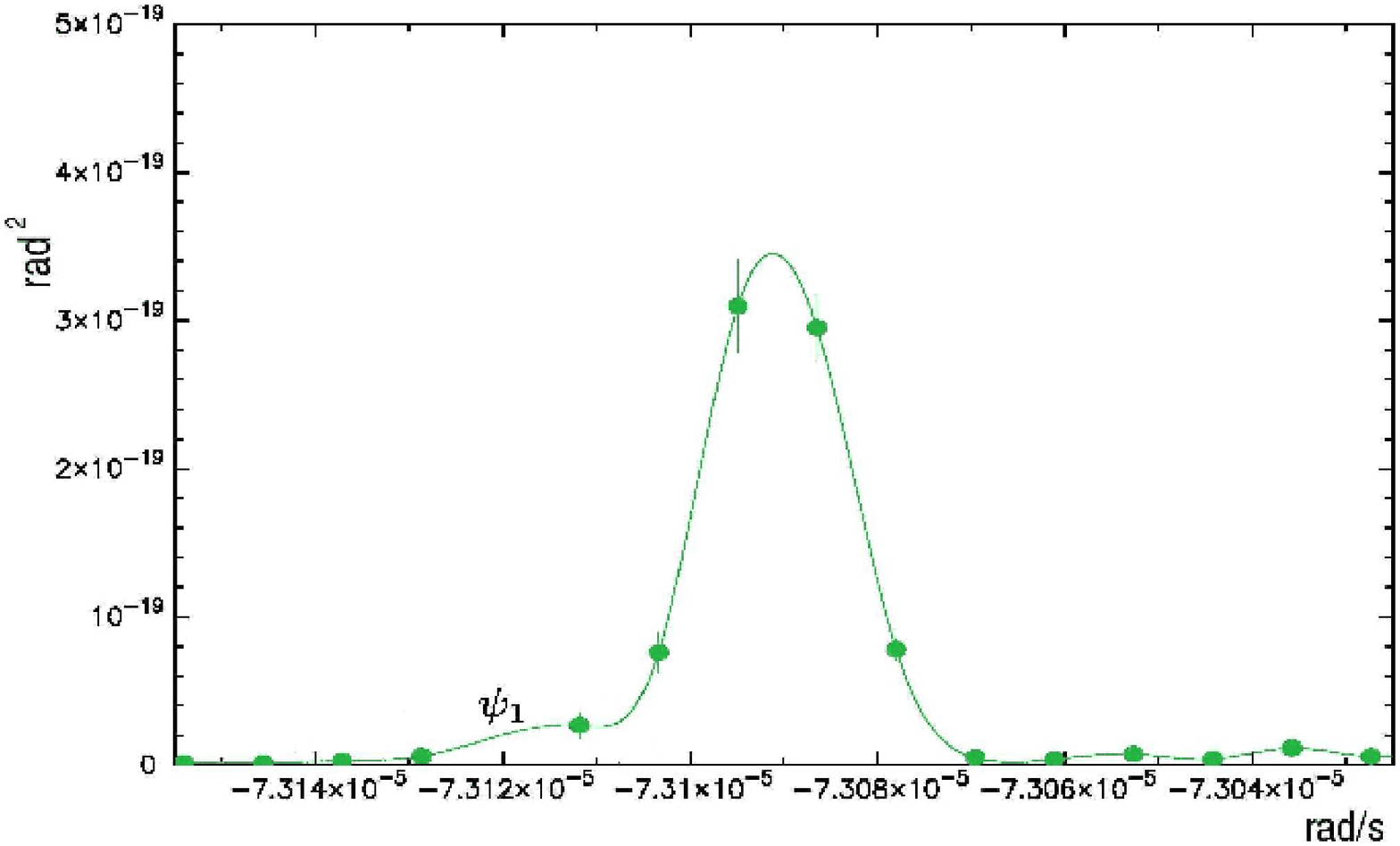}
      \includegraphics[width=0.48\textwidth]{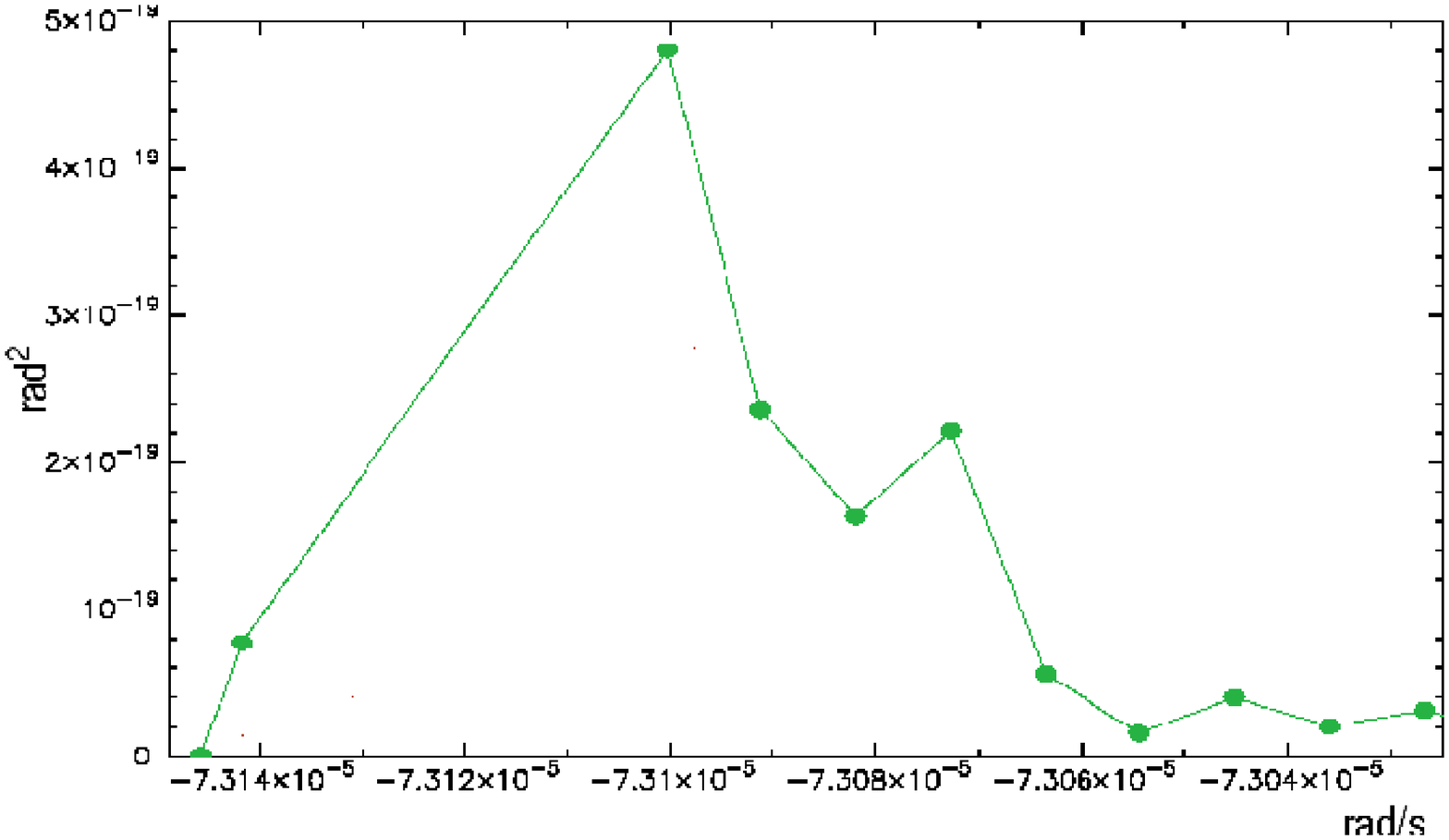}
   \end{center}
   \par\noindent\vspace{-8ex}\par\hphantom{a}
\end{figure}

  Figure~\ref{f:f2} shows an example of two estimates of the RFCN spectrum
made under different assumptions for constituents separation. The estimation 
technique is described in details in \citep{r:pet07}. It should be stressed 
that both spectra equally fit to VLBI group delays. The second spectrum 
shows a bi-modal pattern. We should resist to a temptation to rush inventing
hypothesis for explaining this pattern.

  In order to assess the sensitivity of the RFCN estimates to errors of
oceanic and atmospheric contributions, I ran a Monte Carlo simulation.
I ran 16 solutions and 1)~added the Gaussian noise to estimates of nutations at 
$K_1$, $S_1$, $P_1$ due to uncertainty of ocean contribution with the
standard deviation 150~prad; 2)~added the Gaussian noise to estimates of 
nutations at $\phi_1$, $\psi_1$ with the standard deviation 250~prad; 
3)~obtained new estimates of the empirical transfer function; 4)~computed the 
new set of constraints on forced nutations. The added noise approximately
corresponds to expected uncertainties of the oceanic and atmospheric 
contribution to nutation at these frequencies. Then I computed the rms of 
estimates of the RFCN spectra among these runs:

\begin{tabular}{ll}
  \hem Uncertainty on the RFCN spectrum at different runs:    & 43 prad; \\
  \hem Formal uncertainty on RFCN from noise in group delays: & 19 prad. \\
\end{tabular}

\par\noindent
   We see that constraint uncertainty is dominating.

\section{\large CONCLUSIONS}

  It was shown that the estimates of the RFCN spectrum depend on the 
mathematical model used for separation of the free and forced nutations.
The estimates are not unique. Separation of nutations results in a constraint 
uncertainty. The constraint uncertainty sets the limit of accuracy of the RFCN 
spectrum estimates and dominates the error budget. Since the RFCN has 
a continuous spectrum within its band, more precise observations will not 
result in improving of the accuracy of the RFCN spectrum determination.

\end{document}